\documentclass[10pt,a4paper,onecolumn]{article}
\usepackage{marginnote}
\usepackage{graphicx}
\usepackage[rgb]{xcolor}
\usepackage{authblk,etoolbox}
\usepackage{titlesec}
\usepackage{calc}
\usepackage{tikz}
\usepackage{hyperref}
\hypersetup{%
    unicode=true,
    pdftitle={cTreeBalls: a fast 3-point correlation function code for clustering measurements},
    pdfauthor={Mario A. Rodriguez-Meza},
    colorlinks=true,
    linkcolor=[rgb]{0.0, 0.5, 1.0},
    citecolor=Blue,
    urlcolor=[rgb]{0.0, 0.5, 1.0},
    breaklinks=true
}

\usepackage{caption}
\usepackage{orcidlink}
\usepackage{tcolorbox}
\usepackage{amssymb,amsmath}
\usepackage{ifxetex,ifluatex}
\usepackage{seqsplit}
\usepackage{xstring}

\usepackage{float}
\let\origfigure\figure
\let\endorigfigure\endfigure
\renewenvironment{figure}[1][2] {
    \expandafter\origfigure\expandafter[H]
} {
    \endorigfigure
}

\newlength{\cslhangindent}
\newlength{\csllabelwidth}
\newenvironment{CSLReferences}[2] 
 {
  \setlength{\parindent}{0pt}
  \ifodd #1 \everypar{\setlength{\hangindent}{\cslhangindent}}\ignorespaces\fi
  \ifnum #2 > 0
  \setlength{\parskip}{#2\baselineskip}
  \fi
 }%
 {}
\usepackage{calc}

\usepackage[top=3.5cm, bottom=3cm, right=1.5cm, left=1.0cm,
            headheight=2.2cm, reversemp, includemp, marginparwidth=4.5cm]{geometry}

\titleformat{\section}
  {\normalfont\sffamily\Large\bfseries}
  {}{0pt}{}
\titleformat{\subsection}
  {\normalfont\sffamily\large\bfseries}
  {}{0pt}{}
\titleformat{\subsubsection}
  {\normalfont\sffamily\bfseries}
  {}{0pt}{}
\titleformat*{\paragraph}
  {\sffamily\normalsize}

\usepackage{fancyhdr}
\makeatletter
\let\ps@plain\ps@fancy
\fancyheadoffset[L]{4.5cm}
\fancyfootoffset[L]{4.5cm}

\definecolor{linky}{rgb}{0.0, 0.5, 1.0}

\newtcolorbox{repobox}
   {colback=red, colframe=red!75!black,
     boxrule=0.5pt, arc=2pt, left=6pt, right=6pt, top=3pt, bottom=3pt}

\newcommand{\ExternalLink}{%
   \tikz[x=1.2ex, y=1.2ex, baseline=-0.05ex]{%
       \begin{scope}[x=1ex, y=1ex]
           \clip (-0.1,-0.1)
               --++ (-0, 1.2)
               --++ (0.6, 0)
               --++ (0, -0.6)
               --++ (0.6, 0)
               --++ (0, -1);
           \path[draw,
               line width = 0.5,
               rounded corners=0.5]
               (0,0) rectangle (1,1);
       \end{scope}
       \path[draw, line width = 0.5] (0.5, 0.5)
           -- (1, 1);
       \path[draw, line width = 0.5] (0.6, 1)
           -- (1, 1) -- (1, 0.6);
       }
   }

\patchcmd{\@maketitle}{center}{flushleft}{}{}
\patchcmd{\@maketitle}{center}{flushleft}{}{}
\patchcmd{\@maketitle}{\LARGE}{\LARGE\sffamily}{}{}
\def\maketitle{{%
  
  \AB@maketitle}}
\renewcommand\AB@affilsepx{ \protect\Affilfont}
\renewcommand\AB@affilnote[1]{{\bfseries #1}\hspace{3pt}}
\renewcommand{\affil}[2][]%
   {\newaffiltrue\let\AB@blk@and\AB@pand
      \if\relax#1\relax\def\AB@note{\AB@thenote}\else\def\AB@note{#1}%
        \setcounter{Maxaffil}{0}\fi
        \begingroup
        \let\href=\href@Orig
        \let\protect\@unexpandable@protect
        \def\thanks{\protect\thanks}\def\footnote{\protect\footnote}%
        \@temptokena=\expandafter{\AB@authors}%
        {\def\\{\protect\\\protect\Affilfont}\xdef\AB@temp{#2}}%
         \xdef\AB@authors{\the\@temptokena\AB@las\AB@au@str
         \protect\\[\affilsep]\protect\Affilfont\AB@temp}%
         \gdef\AB@las{}\gdef\AB@au@str{}%
        {\def\\{, \ignorespaces}\xdef\AB@temp{#2}}%
        \@temptokena=\expandafter{\AB@affillist}%
        \xdef\AB@affillist{\the\@temptokena \AB@affilsep
          \AB@affilnote{\AB@note}\protect\Affilfont\AB@temp}%
      \endgroup
       \let\AB@affilsep\AB@affilsepx
}
\makeatother

\renewcommand\Affilfont{\sffamily\small\mdseries}
\ifnum 0\ifxetex 1\fi\ifluatex 1\fi=0 
  \usepackage[T1]{fontenc}
  \usepackage[utf8]{inputenc}

\else 
  \ifxetex
    \usepackage{mathspec}
    \usepackage{fontspec}

  \else
    \usepackage{fontspec}
  \fi
  \defaultfontfeatures{Scale=MatchLowercase}
  \defaultfontfeatures[\sffamily]{Ligatures=TeX}
  \defaultfontfeatures[\rmfamily]{Ligatures=TeX,Scale=1}

\fi
\IfFileExists{upquote.sty}{\usepackage{upquote}}{}
\IfFileExists{microtype.sty}{%
\usepackage{microtype}
\UseMicrotypeSet[protrusion]{basicmath} 
}{}

\PassOptionsToPackage{usenames,dvipsnames}{color} 
\ifLuaTeX
\usepackage[bidi=basic]{babel}
\else
\usepackage[bidi=default]{babel}
\fi
\babelprovide[main,import]{american}

\def\languageshorthands#1{}
\usepackage{longtable,booktabs,array}

\IfFileExists{parskip.sty}{%
\usepackage{parskip}
}{
\setlength{\parindent}{0pt}
\setlength{\parskip}{6pt plus 2pt minus 1pt}
}
\ifx\paragraph\undefined\else
\let\oldparagraph\paragraph
\renewcommand{\paragraph}[1]{\oldparagraph{#1}\mbox{}}
\fi
\ifx\subparagraph\undefined\else
\let\oldsubparagraph\subparagraph
\renewcommand{\subparagraph}[1]{\oldsubparagraph{#1}\mbox{}}
\fi
\ifLuaTeX
  \usepackage{selnolig}  
\fi

\title{cTreeBalls: a fast 3-point correlation function code for clustering measurements}

\author[1%
]
{Mario A. Rodriguez-Meza%
  \,\orcidlink{0000-0003-1160-1488}\,%
}
\author[2%
\ensuremath\mathparagraph
]
{Eladio Moreno%
  \,\orcidlink{0000-0002-5400-2584}\,
}
\author[3%
]
{Alejandro Aviles%
  \,\orcidlink{0000-0001-5998-3986}\,
}
\author[2%
]
{Gustavo Niz%
  \,\orcidlink{0000-0002-1544-8946}\,
}

\affil[1]{Departamento de F\'isica, Instituto Nacional de Investigaciones Nucleares,
Apartado Postal 18-1027, Col. Escand\'on, Ciudad de M\'exico,11801, M\'exico}
\affil[2]{Departamento de Ciencias e Ingenier\'ias, Universidad de Guanajuato,
37150, Le\'on, Guanajuato, M\'exico}
\affil[3]{Instituto de Ciencias F\'isicas, Universidad Nacional Aut\'onoma de M\'exico,
62210, Cuernavaca, Morelos, M\'exico}

\affil[$\mathparagraph$]{Corresponding author}
\date{\vspace{-2.5ex}}

\begin{document}
\maketitle

\marginpar{

  \begin{flushleft}
  \sffamily\small

  {\bfseries DOI:} \href{https://doi.org/xx.xxxxx/joss.xxxxx}{\color{linky}{xx.xxxxx/xxxx.xxxxx}}

  \vspace{2mm}
    {\bfseries Software}
  \begin{itemize}
    \setlength\itemsep{0em}
    \item \href{https://github.com/***jounals/***reviews/issues/5571}{\color{linky}{Review}} \ExternalLink
    \item \href{https://github.com/rodriguezmeza/cTreeBalls}{\color{linky}{Repository}} \ExternalLink
    \item \href{https://doi.org/10.5281/zenodo.10072128}{\color{linky}{Archive}} \ExternalLink
  \end{itemize}

  \vspace{2mm}
  
    \par\noindent\hrulefill\par

  \vspace{2mm}

  {\bfseries Editor:} \href{https://github.com/openjournals}{@openjournals} \\
  \vspace{1mm}
    {\bfseries Reviewers:}
  \begin{itemize}
  \setlength\itemsep{0em}
    \item \href{https://github.com/openjournals}{@openjournals}
    \end{itemize}
    \vspace{2mm}
  
    {\bfseries Submitted:} 9 april 2026\\
    {\bfseries Published:} 

  \vspace{2mm}
  {\bfseries License}\\
  Authors of papers retain copyright and release the work under a Creative Commons Attribution 4.0 International License (\href{https://creativecommons.org/licenses/by/4.0/}{\color{linky}{CC BY 4.0}}).

  \end{flushleft}
}

\hypertarget{summary}{%
\section{Summary}\label{summary}}

\texttt{cTreeBalls}\footnote{\href{http://github.com/rodriguezmeza/cTreeBalls.git/}{http://github.com/rodriguezmeza/cTreeBalls.git/}}
(\texttt{cBalls} for short)
 is a Python/C package useful to measure (2,3)-point clustering statistics.
\texttt{cBalls} can efficiently calculate 3 point correlations of more than 
200 million of healpix pixels
(a full sky simulation with Nside=4096) in  less than 10 minutes
 on a single high-performance computing
node, enabling a feasible analysis for the upcoming LSST data.
It builds upon octree
(\protect\hyperlink{ref-Barnes:1986}{Barnes \& Hut, 1986})
and kd-tree algorithms 
(\protect\hyperlink{ref-Bentley:1975}{Beantly, 1975})
%
and supplies a user-friendly interface with flexible input/output (I/O) of
catalogue data and measurement results, with the built program
configurable through external parameter files and tracked through enhanced
logging and warning/exception handling. For completeness and
complementarity, methods for measuring two-point clustering statistics 
for periodic boxes are also included in the package. 
\texttt{cTreeBalls} was developed for its use in the Dark Energy Science Collaboration (DESC) of the Rubin Observatory Legacy Survey of Space and Time
(LSST).\footnote{\href{http://www.lsst.org/}{http://www.lsst.org}}

\hypertarget{statement-of-need}{%
\section{Statement of need}\label{statement-of-need}}

Correlation functions are frequently used to extract relevant information about the large scale structure of the universe, which in turn can be used to discriminate between cosmological models. A common practice is to employ two point statistics, such as the two point correlation function (2PCF) or its Fourier space counterpart, the power spectrum. A random gaussian field is fully characterized by the two point statistics, however, the non-linear cosmic evolution of matter produces a measurable deviation from this random gaussian field assumption. Equivalently, primordial non-gaussianities may also imprint a non-gaussian signal. Therefore, it is crucial to use alternative methods to extract the non-gaussian information that is not captured by the two point statistics. Higher order point correlations are frequently used as statistical tools to extract such information. The three point correlation function (3PCF), or the bispectrum in Fourier space, is the next order term in the hierarchy of n-point correlation functions based on signal to noise ratio on sufficiently large scales. Just as the 2PCF describes characteristic length-scales of data, the 3PCF encodes all the 
spatial
triangles in that distribution. As a result, an isotropic 3PCF depends on three scales that determine the side lengths of these triangles.

In this work, we provide an infrastructure to calculate 2PCF and 3PCF 
for scalar quantities on the sphere, such as the weak lensing convergence field or the CMB lensing field.
A na\"{\i}ve estimator of the 3PCF requires $O(N^3)$ evaluations, where $N$ is the number of discrete data points or pixels on a continuum field. For present galaxy surveys, such as the 
one that will be released by the LSST
(\protect\hyperlink{ref-Vera:2019}{Ivezi\'c et al., 2019}),
this scaling results in a prohibited amount of time. Using tree structures in the search for neighboring data points can reduce the scaling to $O(N^2\log N)$, which is still not viable. However, one can map the information of the 3PCF onto an orthogonal basis whose coefficients scale as pair counts, $O(N\log N)$, hoping that only a small and finite number of eigenvectors on that basis capture most of the information encoded in the 3PCF. This is the case of a harmonic decomposition of one of the angles in the triangle configurations of the 3PCF for the weak lensing convergence  (\protect\hyperlink{ref-Arvizu:2025}{Arvizu et al., 2025}). 
This expansion was first proposed by Szapudi 
(\protect\hyperlink{ref-Szapudi:2004}{Szapudi, 2004})
and has since been applied to different algorithms and codes 
(e.g. \protect\hyperlink{ref-Zheng:2004}{Zheng, 2004},
\protect\hyperlink{ref-Slepian:2018}{Slepian \& Einsenstein, 2018},
\protect\hyperlink{ref-Philcox:2022}{Philcox \& Slepian 2022}).

Following above ideas
we have developed a code that can efficiently calculate 3 point correlations of more than 200 million pixels
(a full sky simulation with Nside=4096) in less than 10 minutes on a single high-performance computing node.  
Therefore we will be able to analyse the upcoming LSST data
that is planned to be of more than 20 billions of objects.

Specifically, \texttt{cBalls} can calculate:

\begin{itemize}
\item
  three-point clustering statistics, namely multipoles of the three-point correlation function (3PCF) in
  configuration space for projected fields;
\item
  two-point clustering statistics, namely the two-point correlation function (2PCF), for
  projected fields or cubic-box simulation mocks
\end{itemize}


There are existing software packages that use different
decomposition of the three-point clustering statistics or that
use different tree schemes, for instance:

\begin{itemize}
\item
  \texttt{TreeCorr}  (\protect\hyperlink{ref-Jarvis:2004}{Jarvis, et al.,
  2004}) computes (2,3)-point correlation function for counts, convergence and shear
  weak lensing fields. It uses a ball-tree scheme.
 
 \item 
 \texttt{Triumvirate}
   (\protect\hyperlink{ref-Wang:2023}{Wang, et al., 2023})
computes three point statistics in Fourier and configuration spaces using a tri-polar spherical harmonic decomposition 
(\protect\hyperlink{ref-Sugiyama:2019}{Sugiyama et al., 2019}).
\end{itemize}

In particular, the last one 
uses a different decomposition of three-point
clustering statistics and may have different constraining power on different cosmological parameters. Therefore, 
\texttt{cBalls} fulfills
complementary needs in current galaxy clustering analyses.

\hypertarget{implementation}{%
\section{Implementation}\label{implementation}}

A binned estimator to study the 3PCF signal of the weak lensing
convergence field was introduced in 
 \protect\hyperlink{ref-Arvizu:2025}{Arvizu et al., 2025}.
 Extract this kind of signal is prohibitively CPU time consuming
 when we are dealing with a vast amount of sources.
 Brute force algorithms scales with the number of galaxies as
 $O(N^3)$. We can speed up if we introduce a tree structure (kd-tree or octree)
 for the spatial part of the catalog map.
 We are using mainly octree structures
  (\protect\hyperlink{ref-Barnes:1986}{Barnes \& Hut, 1986}), 
 but kd-tree is also used
  (\protect\hyperlink{ref-Bentley:1975}{Bentley, 1975}).
 Now, the real improvement comes when we use a harmonic decomposition
 of the  weak lensing convergence field  
 and impose homogeneity and isotropy, and using the fact the 
  field is a spin-0 field (a real number). 
  In such a case complexity of the code goes as $O(N \log N)$ 
  as it does a 2PCF code. Multipoles are the estimators we are looking for
  that characterize the three-point statistics.

Then, computation loop goes as: pick up a galaxy from the survey catalog,
this is called the pivot, a vertex of a triangle,
 find all its neighbours by scanning the tree in the spatial range of search and fill the multipoles bins.
 Here, we need to evaluate $\sin$ and $\cos$ functions, we instead turn this into 
 evaluation of Chebyshev's polynomials which is much faster than evaluating trigonometric functions.

Another speed improvement comes from the following idea:
spatial closer pivots share neighbor list. We just need to define and quantify ``closeness''. At least there are four length scales. First, the length scale of the root in the octree hierarchy, the cube that contain all the galaxies to be analyzed. Second, the
scale length of the smallest cell in this hierarchy. Third, the radial length of each bin in the histograms that contain all correlation information. And fourth, the mean separation between galaxies. \texttt{cBalls} can compute the pivot radius of the neighborhood automatically using the latter scale length. But user can give one if needed.

Furthermore, octree cells can be tightened (by pruning) and create a kind of \emph{oc-ball-tree}.
Therefore, when scanning the tree, we only visit regions which are filled up with galaxies,
instead of visiting cubic cells which may have not negligible void space.

\hypertarget{features}{%
\section{Features}\label{features}}

\texttt{cBalls} has many
features that makes it appropriate to be integrated
in massive workflow like \texttt{TXPipe} 
(\protect\hyperlink{ref-Prat:2023}{Prat et al., 2023})
used in
Vera C. Rubin Observatory with its Legacy Survey of Space and Time
(LSST) collaboration\footnote{\href{http://www.lsst.org/}{http://www.lsst.org}}:

\begin{itemize}
\item
  The code has a Python interface 
  (\texttt{cyballs})
  that was implemented
  using Cython to bind the C searching engine. Python scripts are given
  in such a way users can run the code in a linux terminal or can be run
  using a Jupyter notebook. The C code can also be compiled and 
  executed independently of Python.
  \item
  Parameters files can be given as standard \texttt{ASCII} files or in command 
  line. If using a Jupyter Notebook or a Python script, parameters are
  given very easily as a very simple dictionary without the need of a
  parameter file. Before computing correlations parameter values are checked
  to secure the computation.
\item
 Catalog map files can be given in several convenient formats, such as
 \texttt{CFITSIO}, \texttt{HEALPix},
 \texttt{GADGET-2} 
 (\protect\hyperlink{ref-Springel:2005}{Springel, 2005}) 
 and  \texttt{ASCII} 
 columns format, like \texttt{ROCKSTAR} 
   (\protect\hyperlink{ref-Behroozi:2013}{Behroozi et al., 2013}) 
  halos catalogs.
  Also, Takahashi \emph{et al}. 
  (\protect\hyperlink{ref-Takahashi:2017}{Takahashi et al., 2017}) 
  weak lensing realizations can be
  read by the code without being transformed to \texttt{HEALPix} format. 
  Several map files can be read in order
  to compute cross-correlations function or to apply a mask file that is
  useful for current surveys that are unable to scan the full sky.
\item
  Searching engines are parallelised with OpenMP threading, i.e.,
  pivot  pixels or galaxies are distributed amongst multiple CPU threads.
\item
  Clustering statistics can be done on a cubic box with periodic boundary
  condition (only 2pcf) or on a unit sphere under the Limber approximation for 
  projected scalar fields.
\item
  Second and third order statistics can be done in a single run
  or they can be done independently.  
 \item
 Edge corrections are implemented
     (\protect\hyperlink{ref-Slepian:2015}{Slepian \& Eisenstein, 2015})
     in the case of computing 3-pcf multipoles over volumes that are not the full sky. 
\item
  A four levels logger is provided for runtime tracking. Running
  information can be printed on the screen or saved in a log file.
  CPU time and memory usage can be reported at each main stages of the computation.
\end{itemize}

\hypertarget{performance}{%
\section{Performance}\label{performance}}

Brute force algorithms to compute 3-point correlation function scale
as $O(N^3)$.
Tree methods can improve algorithms performance. 
However, as explained above, CPU time can be dramatically reduced if we use
a harmonic basis decomposition, a sharing neighbor list and a tighter oc-ball-tree structure. 
Resulting that computing this
higher order statistics will scale with the number of objects 
similar to computing 2-point correlation function. Full sky analysis for an Nside$\,=8192$, around
800 millions of pixels and for maximal searching radius distance of 200 arcmin takes around of
40 CPU minutes (wall-clock) on 128 threads of a single Perlmutter-NERSC 
node.\footnote{\href{https:/docs.nersc.gov/systems/perlmutter/architecture}{https:/docs.nersc.gov/systems/perlmutter/architecture}}
For this vast amount of pixels the CPU time consumption comes mainly from scanning the three and updating the histograms by summing up the neighbor lists. For each multipole component, we need to compute its corresponding Chebyshev polynomials which translate toa na\"ive $O(m_{max}N\log N)$ scaling, where $m_{max}$ is the maximal multipole number. In summary, we found 
(\protect\hyperlink{ref-Arvizu:2025}{Arvizu et al., 2025}), 
for a set of \texttt{HEALPix} Nsides parameters 256, 512, 1024, 2048 and 4096, that the power scaling law goes as $N^{1.1}$. The scaling with maximum number of multipoles shows a linear scaling $t(m)=1.85m+63.7$ seconds for Nside $=1024$. We notice that the slope of $t(m)$ is close to 2, which is the number of operations needed in the recursive relations of the Chebyshev polynomials. However, the dependence on $m$ is moderate, being $\sim 20$ \% slower to compute up to $m=8$ than up to $m=1$.

\hypertarget{acknowledgements}{%
\section{Acknowledgements}\label{acknowledgements}}

The authors acknowledge support by SECIHTI (previously CONAHCyT) grants CBF2023-2024-162,  CBF2023-2024-589 and BF-2025-I-2795. AA also acknowledges DGAPA-PAPIIT IA101825. EM and GN acknowledge the support of the DAIP-UG grant CIIC-254/2026 and the computational resources of the DCI-UG DataLab.
MARM also acknowledges 
the computational resources of the 
DiRAC@Durham facility managed by the Institute for Computational Cosmology on behalf 
of the STFC DiRAC HPC Facility (www.dirac.ac.uk). 

For the purpose of open access, the authors have applied a
Creative Commons Attribution (CC BY) license to any Author Accepted
Manuscript version arising from this submission. 

\hypertarget{references}{%
\section*{References}\label{references}}
\addcontentsline{toc}{section}{References}

\hypertarget{refs}{}
\begin{CSLReferences}{1}{0}

\leavevmode\vadjust pre{\hypertarget{ref-Arvizu:2025}{}}%
Arvizu, A., Aviles, A., Hidalgo, J.C., Moreno, E.,
Niz, G., Rodriguez-Meza, M.A. \& Samario, S.,
LSST Dark Energy Science collaboration. 
(2025).
\emph{JCAP}, \emph{12}(2024), 049.
\url{https://doi.org/10.1088/1475-7516/2024/12/049}

\leavevmode\vadjust pre{\hypertarget{ref-Barnes:1986}{}}%
Barnes, J. \& Hut, P. (1986). 
\emph{A hierarchical $O(N log N)$ force-calculation algorithm}.
\emph{Nature}, \emph{324}(December), 446--449.

\leavevmode\vadjust pre{\hypertarget{ref-Bentley:1975}{}}%
Bentley, J.L. (1975).
\emph{Multidimensional binary search trees used for associative searching}.
\emph{Commun. ACM}, \emph{18}(September), 509--517.

\leavevmode\vadjust pre{\hypertarget{ref-Behroozi:2013}{}}%
Behroozi, P.S., Wechsler, R.H., \& Wu, H.-Y. (2013).
\emph{Astrophys.~J.}, 
\emph{762}(109), 20pp. \url{https://doi.org/10.1088/0004-637X/762/2/109}

\leavevmode\vadjust pre{\hypertarget{ref-Vera:2019}{}}%
Ivezi\'c, Z., et al. (2019).
LSST: From Science Drivers to Reference Design and Anticipated Data Products.
\emph{Astrophys.~J.}, 
\emph{873}(2), 44pp. \url{https://doi.org/10.3847/1538-4357/ab042c}

\leavevmode\vadjust pre{\hypertarget{ref-Jarvis:2004}{}}%
Jarvis, M., Bernstein, G.,  \& Jain, B. (2004).
\emph{Mon.~Not.~R.~Astron.~Soc.}, 
\emph{352}, 338. \url{https://doi.org/ 10.1111/j.1365-2966.2004.07926.x}

\leavevmode\vadjust pre{\hypertarget{ref-Philcox:2022}{}}%
Philcox, O.H.E. \& Slepian, Z. (2022).
Efficient computation of N-point correlation functions in D
dimensions.
\emph{Proc.~Nat.~Acad.~Sci.}, 
\emph{119}(33), e2111366119. 
\url{https://doi.org/10.1073/pnas.21113661}

\leavevmode\vadjust pre{\hypertarget{ref-Prat:2023}{}}%
Prat, J. and Zuntz, J. and Omori, Y. and Chang, C. and Tr{\" o}ster, T. and Pedersen, E. and Garc{\' i}a-Garc{\' i}a, C. and Phillips-Longley, E. and Sanchez, J. and Alonso, D. and Fang, X. and Gawiser, E. and Ishak, M. and Heitmann, K. and Jarvis, M. and Kovacs, E. and Mao, Y. -Y. and Larsen, P. and Varela, L. Medina and Paterno, M. and Vitenti, S. D. and Zhang, Z. and Collaboration, The LSST Dark Energy Science. (2023).
The catalog-to-cosmology framework for weak lensing and galaxy clustering for {LSST}.
The Open Journal of Astrophysics \emph{6}(April).
\url{https://doi.org/10.21105/astro.2212.09345}

\leavevmode\vadjust pre{\hypertarget{ref-Slepian:2015}{}}%
Slepian, Z., \& Eisenstein, D. J. (2015). Computing the three-point
correlation function of galaxies in \(\mathcal{O}(N^2)\) time.
\emph{Mon.~Not.~R.~Astron.~Soc.}, \emph{454}(4), 4142–4158.
\url{https://doi.org/10.1093/mnras/stv2119}

\leavevmode\vadjust pre{\hypertarget{ref-Slepian:2018}{}}%
Slepian, Z. \& Eisenstein, D.J. (2018).
A practical computational method for the anisotropic
redshift-space three-point correlation function.
\emph{Mon.~Not.~R.~Astron.~Soc.}, 
\emph{469}(2), 1468--1483. 
\url{https://doi.org/10.1093/mnras/sty1063}

\leavevmode\vadjust pre{\hypertarget{ref-Springel:2005}{}}%
Springel, V. (2005). The cosmological simulation code GADGET-2.
\emph{Mon.~Not.~R.~Astron.~Soc.}, \emph{364}, 1105–1134.
\url{https://doi.org/10.1111/j.1365-2966.2005.09655.x}

\leavevmode\vadjust pre{\hypertarget{ref-Sugiyama:2019}{}}%
Sugiyama, N. S., Saito, S., Beutler, F., \& Seo, H.-J. (2019). A
complete {FFT}-based decomposition formalism for the redshift-space
bispectrum. \emph{Mon.~Not.~R.~Astron.~Soc.}, \emph{484}(1), 364–384.
\url{https://doi.org/10.1093/mnras/sty3249}

\leavevmode\vadjust pre{\hypertarget{ref-Szapudi:2004}{}}%
Szapudi, I. (2004).
Three-point statistics from a new perspective.
\emph{Astrophys.~J. Lett.}, 
\emph{605}(L89), 527. 
\url{https://doi.org/10.1086/420894}

\leavevmode\vadjust pre{\hypertarget{ref-Takahashi:2017}{}}%
Takahashi, R., Hamana, T., Shirasaki, M., Namikawa, T., 
Nishimichi, T., Osato, K., \& Shiroyama, K. (2017).
\emph{Astrophys.~J.}, 
\emph{850}(24), 23pp. \url{https://doi.org/10.3847/1538-4357/aa943d}

\leavevmode\vadjust pre{\hypertarget{ref-Wang:2023}{}}%
Wang, M.S., Beutler, F.,  \& Naonori, S. (2023).
\emph{Journal of Open Source Software}, 
\emph{8}(91), 6pp. \url{https://doi.org/ 10.21105/joss.05571}

\leavevmode\vadjust pre{\hypertarget{ref-Zheng:2004}{}}%
Zheng, Z. (2004).
Projected three-point correlation functions and galaxy bias.
\emph{Astrophys.~J.}, 
\emph{614}, 527--532. 
\url{https://doi.org/10.1086/423838}

\end{CSLReferences}

\end{document}